# Implications of Environmental Uncertainty for Business-IT Alignment: A Comparative Study of SMEs and Large Organizations

**Amitha Padukkage**
School of Information Management
Victoria University of Wellington
Wellington, New Zealand
Email: amitha.padukkage@vuw.ac.nz

**Val Hooper**
School of Information Management
Victoria University of Wellington
Wellington, New Zealand
Email: val.hooper@vuw.ac.nz

**Janet Toland**
School of Information Management
Victoria University of Wellington
Wellington, New Zealand
Email: janet.toland@vuw.ac.nz

## Abstract

This paper presents a comprehensive study of the influence of environmental uncertainty on business-IT alignment. The existing literature postulates environmental uncertainty as a key challenge to achieving business-IT alignment. Hence, the first objective of this study is to identify the extent of the impact of environmental uncertainty on business-IT alignment, and to determine its relative impact in the light of the other antecedents. Furthermore, small and medium sized enterprises (SMEs) differ fundamentally from large firms in many ways. Thus this paper also aims to investigate the variation between SMEs and large firms with regard to the antecedents for strategic alignment. Based on data collected from 212 firms, a conceptual model is tested against the research objectives. The findings provide important contributions to both research and practice by demonstrating the relative impact of environmental uncertainty, and showing how the antecedents of alignment vary between SMEs and large firms.

## 1 Introduction

Aligning information technology (IT) strategy with business strategy is viewed as an important driver for optimizing business performance and it remains one of the top concerns for IT and business executives (Chan et al. 2007; Luftman et al. 2012). Alignment is a collaborative process between business and IT managers which enables them to search for opportunities for embedding IT into business (Choe 2003). Different perspectives have been used to define alignment (Reich et al. 1996); to achieve alignment (Chan et al. 2006; Hu et al. 2006; Reich et al. 2000); and to explore its impact on organizational performance (Chan et al. 2006; Yayla et al. 2012). Researchers have identified numerous factors that contribute to strategic alignment but key factors are shared domain knowledge, prior IS successes, relationship management, communication, and planning connection (Chan et al. 2006; Hu et al. 2006; Luftman et al. 1999; Reich et al. 2000). Despite the identification of these factors, attaining alignment between IT and business remains a challenge (Luftman et al. 2012). Environmental uncertainty, in particular, has been found to be the one of the key challenges to achieving strategic alignment (Sabherwal et al. 2001; Sabherwal et al. 1994).

Environmental uncertainty refers to the perceived unpredictability of environmental variables that have an impact on an organization's performance (Miller 1993). It is often caused by changes in markets, technologies and the regulatory environment (Bstieler 2005; Engau et al. 2009). Uncertainty increases the difficulties in understanding the environment and places managers in a challenging situation with regards to strategic decision making (Johnson et al. 2005; Xu et al. 2005). Required information for making comprehensive decisions may not be available (Fredrickson et al. 1984). This lack of information together with the unpredictability of environmental variables may lead to serious mistakes in decision making (Johnson et al. 2005; Johnston et al. 2008; Xu et al. 2005) and may inhibit coordination between IT and business units and lead to conflicting unit goals (Sabherwal et al. 1994). An uncertain environment may therefore be associated with a lower level of business-IT alignment (Sabherwal et al. 1994). Even after an organization has achieved alignment, the environment continues to change, whether slowly or rapidly. Thus, an organization needs to continually fine-tune their alignment to accommodate ongoing environmental changes, making it important for managers to develop an understanding of the effect of environmental uncertainty on strategic alignment.

Therefore the first objective of this paper is to identify the extent of the impact of environmental uncertainty on the alignment of business and IT, and to determine its relative impact in light of the other antecedents. To pursue this objective, a conceptual model of strategic alignment and its antecedents was developed and empirically tested in the context of a developing country. The environment in developing countries is considered less stable and is often marked by strong turbulence (Iakovleva et al. 2011). Thus, organizations in these countries generally experience higher environmental uncertainty than organizations in developed countries (Iakovleva et al. 2011).

Furthermore, organizations in developing countries consist of predominantly more SMEs than in developed countries (Abor et al. 2010). SMEs differ fundamentally from large firms in many ways, and these differences can affect the way they deal with environmental uncertainty (Johnston et al. 2008). Their interpretation of environmental uncertainty is different (Lester et al. 2007), and whereas large firms base decisions about current strategic planning on years of accumulated data and experience, involving multiple stakeholders, SMEs frequently rely on one or a few owners/managers who tend to make more intuitive decisions (Parnell et al. 2012). This makes it appropriate to analyse SMEs and large firms separately.

In addition, most of the proposed strategic alignment models and recommendations are based on studies of large firms (Dwivedi et al. 2009). Yet firm size has been found to affect strategic alignment (Chan et al., 2006). SMEs are more likely to have centralized structures with centrally coordinated functions. This central coordination limits the need for other explicit mechanisms to promote functional alignment (Chan et al. 2006). On the other hand, large organizations generally use decentralized governance structures that make coordination more difficult. They need more resources and mechanisms to promote coordination and thus to promote strategic alignment (Chan et al. 2006; Dwivedi et al. 2009). These differences are likely to influence the way SMEs approach aligning IT with business strategy. Therefore the second objective of this research is to identify whether the antecedents of strategic alignment manifest different effects in SMEs and large firms.

This paper is organized in the following way. The conceptual model is described in the next section. Section 3 describes the research process and the methods employed. Data analysis is presented in section 4, and Section 5 presents the discussion of the results. Conclusions and implications of this research are discussed in section 6.

## 2  Conceptual model

Strategic alignment has two dimensions. The intellectual dimension concentrates on examining the strategies, structure, and planning methodologies in organizations (Chan et al., 1997). The social dimension investigates the actors in organizations, examining their values, communication with each other, and ultimately their understanding of each other's domains (Reich & Benbasat, 2000). Similarly, strategic alignment can be approached from a process or a state perspective (Reich et al. 2000). A process perspective is concerned with planned activities which are performed dynamically through the iterative process of achieving alignment (Gutierrez et al. 2008). The state perspective views alignment as a fixed outcome which can be achieved by the manipulation of a number of antecedents (Reich et al. 2000). This study focus on the social dimension and, strategic alignment refers to "the extent to which the IT mission, objectives, and plans support and are supported by the business mission, objectives, and plans" (Reich and Benbasat, 2000, p. 82). In this context strategic alignment is treated as a point or state which is achieved by manipulating a number of antecedents, i.e. during the strategic alignment process.

This paper examines the effects of four antecedents on strategic alignment, three of which – shared domain knowledge, relationship management and prior IS success – are adapted from prior empirical research on strategic alignment (Hu et al. 2006; Reich et al. 2000; Yayla et al. 2009). Shared domain knowledge, prior IS success, and relationship management have been identified as key antecedents of strategic alignment (Hu et al. 2006; Reich et al. 1996; Reich et al. 2000; Yayla et al. 2009). These three antecedents facilitate managerial practices and thus indirectly influence strategic alignment (Hu et al. 2006; Reich et al. 2000; Yayla et al. 2009). Managerial practices that are particularly relevant are communication between business and IT executives, and the connection between business and IT planning processes (Hu et al. 2006; Reich et al. 1996; Reich et al. 2000; Yayla et al. 2009). All these three factors are internal to the organization.

The fourth factor, environmental uncertainty, is proposed as an external antecedent to strategic alignment. Business executives view the influence of uncertainty as one of the most difficult aspects of IS strategic planning (Lederer et al. 1986). It works as an inhibitor of business-IT alignment (Sabherwal and Kirs 1994) and can cause managers to adapt new strategies and tactics more frequently (Bstieler 2005; Calantone et al. 2003). In all, this frequent communication between executives and planning process plays a critical role. Based on these arguments this research proposes that environmental uncertainty will influence both communication between business and IT executives and connections between business and IT planning connection. For more extensive explanation of the model development, see Padukkage et al. (2014). Figure 1. illustrates the proposed research model and the list of hypotheses follows below.

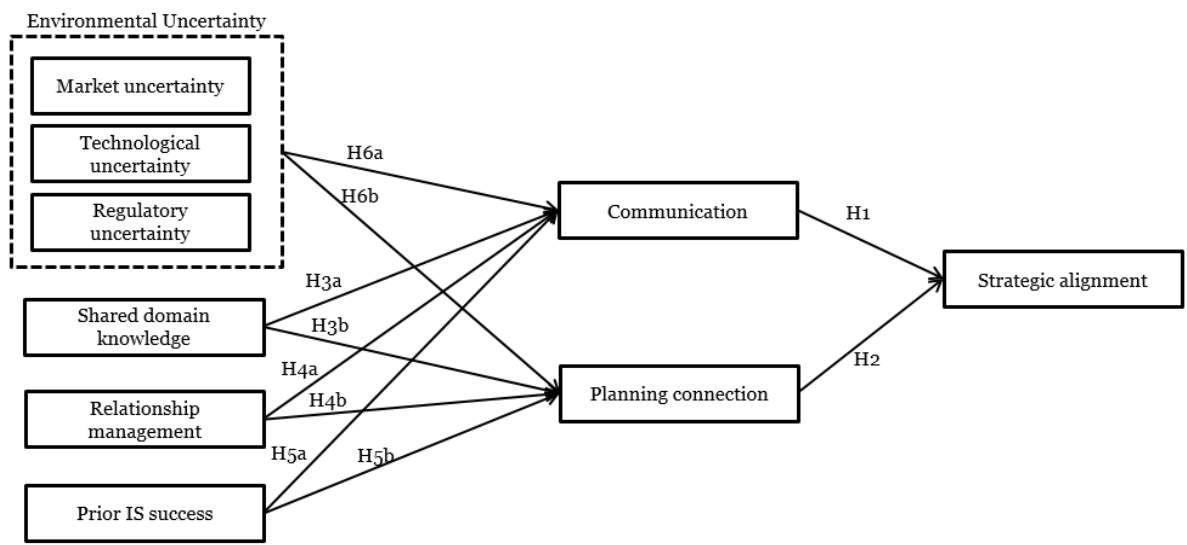

*Figure 1. Conceptual Model*

H1: Communication between business and IT executives will influence the strategic alignment between business and IT

H2: Planning connection between business and IT planning processes will influence the strategic alignment between business and IT

H3a: Shared business and IT domain knowledge will positively influence communication between business and IT executives

H3b: Shared business and IT domain knowledge will positively influence the connections between business and IT planning processes

H4a: Relationship management will influence the communication between business and IT executives

H4b: Relationship management will influence the connections between the business and IT planning processes

H5a: Prior IS success will positively influence the communication between business and IT executives

H5b: Prior IS success will positively influence the connections between business and IT planning processes

H6a: Environmental uncertainty will influence communication between business and IT executives

H6b: Environmental uncertainty will influence connection between business and IT planning processes

## 3   Data collection

The respective literature for each construct was reviewed in order to generate the required items for the questionnaire. Suitable items, which had been validated according to their respective constructs in previous research, were selected from the literature and adapted as necessary. A five-point Likert scale was used for all measurement items. In addition to the measurement items, questions covering demographics of the respondents and participating firms were also included. (The list of items and their sources is available from the authors.) A pilot study was conducted in order to identify items that may be problematic and to ensure that the instrument had acceptable measurement properties. Thirteen CEOs in different industries participated. The results were satisfactory and the questionnaire did not require any significant changes, apart from the addition of two items to the shared domain knowledge construct.

For the actual survey, all firms listed in the directory of the Chamber of Commerce in Sri Lanka and the Registry of Government owned firms in Sri Lanka were grouped under the Central Bank of Sri Lanka Industrial Classification 2012 codes. A stratified sampling technique according to industry sector was used to select a sample of 720 firms. This ensured each of the industry categories was represented proportionally within the sample. Firms in each category were selected using a systematic sampling technique.

In each firm, the Chief Executive Officer (CEO) was asked to participate in the survey, since the CEO is likely to be involved in the strategic planning of the firm, and have a holistic view of the organization's activities. It was important that the CEOs had been in an organization long enough to have had sufficient knowledge of its strategic planning process. Thus, only CEOs who had a minimum of two years' experience were considered for the data analysis. Initial contact to recruit CEOs was made through telephone calls. The questionnaires were mailed to the 720 CEOs who agreed to participate in the survey. A total of 212 respondents returned fully completed questionnaires giving a response rate of 29%. All 212 responses were considered for the analysis.

There are many different ways of defining SMEs, varying from the number of employees to revenue and other variables (e.g. balance sheet total). However a major variable that is repeatedly mentioned in the literature is the number of employees. The World Bank's definition of SME in South Asian countries, suggests small firms have 1 to 50 employees, medium firms 51-300, and large firms 300 and above (UNDP 2012). This scale has been used to distinguish between SMEs and large firms in this paper. Table 1 shows the summary of the participating firms.

| Industry sector | Response information | | | |
|---|---|---|---|---|
|  | SME | | Large | |
|  | No. | % | No. | % |
| Agriculture, Livestock and Forestry | 6 | 6.7% | 12 | 9.76% |
| Banking, Insurance and Real Estate | 16 | 18.0% | 24 | 19.51% |
| Construction | 2 | 2.2% | 4 | 3.25% |
| Fishing | 0 | 0.0% | 1 | 0.81% |
| Electricity, Gas and Water | 7 | 7.9% | 1 | 0.81% |
| Government Services | 12 | 13.5% | 17 | 13.82% |
| Hotels and Restaurants | 6 | 6.7% | 9 | 7.32% |
| Manufacturing | 17 | 19.1% | 36 | 29.27% |
| Mining and Quarrying | 2 | 2.2% | 1 | 0.81% |
| Ownership of Dwellings | 0 | 0.0% | 0 | 0.00% |
| Private Services | 12 | 13.5% | 8 | 6.50% |
| Transport and Communication | 6 | 6.7% | 5 | 4.07% |
| Wholesale and Retail Trade | 3 | 3.4% | 5 | 4.07% |
| Total | 89 | 100.0% | 123 | 100.00% |

*Table 1. Composition of the Sample*

## 4   Data analysis

As this research used a self-reporting survey, there was a possibility of the problem of common method variance, so Harman's one factor test was applied to test for the effects of such variance (Igbaria, Zinatelli, Cragg, & Cavaye, 1997; Podsakoff et al., 2003). The exploratory factor analysis (EFA) with unrotated solution resulted in 10 factors with eigenvalues higher than one, and the most covariance explained by one factor was 27%. Thus no single factor emerged to account for the majority of the variance, so common method variance was excluded as a threat to the validity of the findings. The questionnaire had also been pretested for ambiguity with no indication of such; the items had come from previously validated measures; the questions were not socially or professionally threatening; and the research was conducted across multiple industries/contexts (Podsakoff et al. 2003).

A time-trend extrapolation test was used to examine the non-response bias. Comparison of the responses of early and late responders is based on the assumption that the respondents who respond late are more likely to resemble non-respondents (Armstrong et al. 1977). Two subsamples consisting of the first and last 50 responses were compared using a two-tailed t-test at 5% significance level (Field, 2010). Of the 50 measurement items, only one item (RU2) presented some degree of statistical difference between the two groups. Thus, the findings are consistent with the absence of non-response bias.

Partial Least Squares structural equation modelling was used to test the conceptual model. Both the measurement model and structural model were estimated using Smart PLS 2.0. The assessment of internal consistency, indicator reliability, and convergent validity is summarized in Appendix 1. The following subsections describe the evaluation of the measurement model and structural model for all firms, and then describe the comparison between SMEs and large firms.

### 4.1 Measurement Model evaluation

In this research, environmental uncertainty was treated as a second order construct and was formed using three first order constructs, market uncertainty, technological uncertainty and regulatory uncertainty. Market uncertainty refers to the unpredictability of markets, changes in market structure and the degree of competition with respect to industry (Bstieler 2005). Technological uncertainty refers to the unpredictability of the complexity, and rapid and significant change of the technology (Bstieler 2005). Regulatory uncertainty refers to the unpredictability of the actions of regulatory agencies which create and enforce regulations and policies (Engau et al. 2009). As suggested by Hair et al. (2014), reflective-reflective type was used to define the environmental uncertainty construct. The "two step approach" was used to validate environmental uncertainty. In the first step the latent variable scores were estimated in the model without the second order construct. In the second step the latent variable scores were used as indicators of the second order construct (Chin 2010).

Outer loadings of the indicators and the AVE were utilized to test for the convergent validity of the whole model. The results indicated that the majority of the items' loadings were above 0.7 and statistically significant at the 0.001 level while only three items (SDK4, PISS3, and RM4) had a loading below 0.70. However, the loadings of all the factors were close to the 0.7 and above the acceptable level of 0.6 (Chin, 1998).

The assessment of the discriminant validity of the constructs led to fully satisfactory results which are presented in Table 2. None of the constructs' cross-correlations (off-diagonal elements) exceeded the respective square root of each construct's AVE (diagonal element). Further, the indicator cross loadings showed that each indicator's loadings with its associated construct was higher than any of its loadings with the other constructs of the model. Thus, both the AVE analysis and the examination of cross loadings provided adequate statistical support for the discriminant validity of the main constructs.

| Construct | COM   | CON   | MU    | PISS  | RM    | RU    | SA    | SDK   | TU    |
|-----------|-------|-------|-------|-------|-------|-------|-------|-------|-------|
| COM       | **0.822** |       |       |       |       |       |       |       |       |
| CON       | 0.575 | **0.850** |       |       |       |       |       |       |       |
| MU        | 0.233 | 0.248 | **0.826** |       |       |       |       |       |       |
| PISS      | 0.502 | 0.558 | 0.272 | **0.777** |       |       |       |       |       |
| RM        | 0.539 | 0.495 | 0.215 | 0.506 | **0.850** |       |       |       |       |
| RU        | 0.215 | 0.258 | 0.400 | 0.380 | 0.298 | **0.802** |       |       |       |
| SA        | 0.523 | 0.590 | 0.289 | 0.521 | 0.397 | 0.229 | **0.804** |       |       |
| SDK       | 0.552 | 0.521 | 0.180 | 0.564 | 0.618 | 0.157 | 0.527 | **0.761** |       |
| TU        | 0.270 | 0.332 | 0.595 | 0.317 | 0.305 | 0.537 | 0.330 | 0.225 | **0.865** |

*Table 2. Correlations between main constructs (Diagonal elements are square root of AVE)*

*COM: Communication, CON: Planning Connection, MU: Market Uncertainty, PISS: Prior IS Success, RM: Relationship Management, RU: Regulatory Uncertainty, SA: Strategic Alignment, SDK: Shared Domain Knowledge and TU: Technological Uncertainty.*

Cronbach's alpha coefficients and composite reliability scores were satisfactory and reliability was demonstrated. As presented in Appendix 1, many Cronbach's alpha scores exceed the 0.70 threshold recommended by Chin et al. (2003). However, the prior IS success construct had a slightly lower Cronbach's alpha score at 0.676. Deletion of items did not increase the Cronbach's alpha and given that the alpha score was not far from the threshold point, it was retained and considered for further analysis.

## 4.2 Structural Model evaluation

The predictive power of the model was assessed using the coefficient of determination ($R^2$) value for the dependent constructs (Chin 1998; Chin et al. 2003; Hair et al. 2014). The degree to which the variance in the dependent variables was explained by the independent variables was determined by the R2 values associated with each dependent construct (Chin 1998; Hair et al. 2014; Hulland 1999). Falk and Miller (1992) and Hair et al. (2014) suggest that the minimum acceptable level for an individual $R^2$ should be greater than 0.10. Further they suggest comparing R-square values to the following benchmark levels: 0.25 (weak), 0.5 (moderate), and 0.75 (substantial).

Stone-Geisser's Q2 value and the global goodness-of-fit (GoF) criterion was used to evaluate the quality of the model. Positive $Q^2$ scores indicate that a model has predictive relevance while a negative $Q^2$ means a lack of predictive relevance (Chin, 2010; Fornell & Cha, 1994; Vinzi et al., 2010). The Stone-Geisser $Q^2$ test was performed through a blindfolding procedure in Smart-PLS for evaluating the predictive relevance of the structural model. GoF is the geometric mean of the average communality (equivalent to AVE in PLS path analysis) and the average $R^2$ of endogenous or dependent variables. GoF is normed between 0 and 1, where a higher value represents better path model estimations. GoF scores of 0.1, 0.25, and 0.36 respectively correspond to small, medium, and large effect sizes.

The structural model results provided support for the conceptual model. Figure 2. provides the R2 and path coefficients of this model. The model explained 39.8%, 40.3% and 40.7% of the variance in strategic alignment, communication and planning connection respectively. The GoF value for the model was found to be large (0.526), indicating strong evidence for the overall quality of the model.

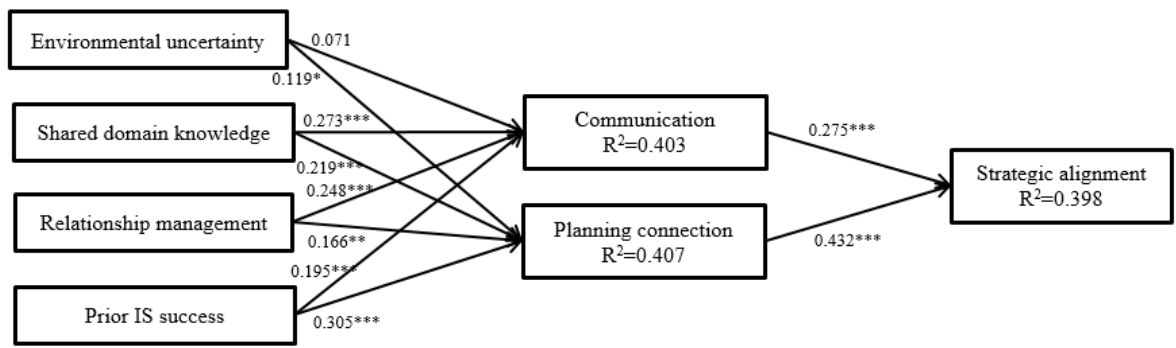

*Figure 2. Summary of the results (\* P<0.10, \*\* p<0.05, \*\*\* p<0.01)*

## 4.3 Comparison between SME and large firms

One of the objectives of this study was to identify the differences between small, medium and large firms regarding the relative impact of antecedents on strategic alignment. Because relevant differences have been found between SMEs and large firms in terms of centralization of structure, degree of centralized governance, and coordination (Gutierrez et al. 2009), small and medium firms were grouped together and the dataset was split into two categories, SMEs and large firms. The evaluation of measurement model and structural model were conducted separately.

The measurement model evaluation indicated that all the items' loadings in the large firms dataset were above the threshold and statistically significant at the 0.001. In the SME dataset, SDK4 was found to be slightly lower at 0.611. However, it was still above the acceptable level of 0.6 suggested by Chin (1998) and thus all items were retained for the analysis.

Discriminant validity for both datasets, SME and large firms, was demonstrated. None of the constructs' cross-correlations exceeded the respective square root of each construct's AVE. Further, the examination of indicator cross loadings also provided adequate statistical support for the discriminant validity of the main constructs for both datasets.

As presented in Appendix 1, many Cronbach's alpha scores exceeded 0.70. The prior IS success constructs had slightly lower Cronbach's alpha scores (SME: 0.648, Large: 0.697). The deletion of no item increased the Cronbach alphas of either construct, and as they were above the acceptable level of 0.6 (Hulland 1999; Wong 2013), all items were retained for further analysis.

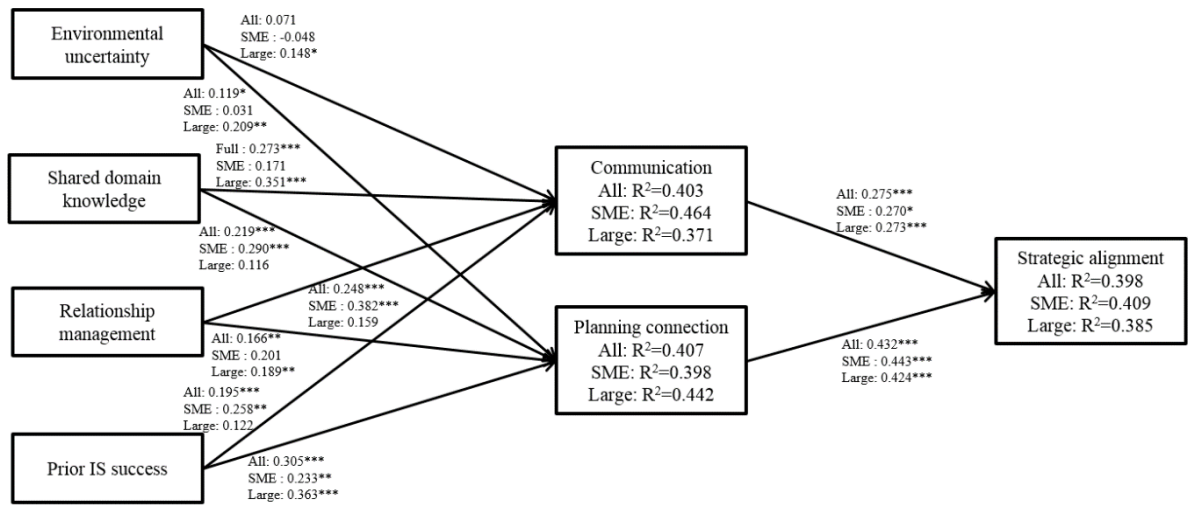

*Figure 3. Summary of the results for all firms, SME and large firms (\* P<0.10, \*\* p<0.05, \*\*\* p<0.01)*

The structural model results provided support for the conceptual model. Figure 3 provides the R2 and path coefficients of this model. Overall, the model explained 39.8%, 40.3% and 40.7% of the variance in strategic alignment, communication and planning connection respectively. In SMEs, the model explained 40.9%, 46.4% and 39.8% of the variance in strategic alignment, communication and planning connection respectively. The corresponding figures for the large firms were 38.5%, 37.1% and 44.2% respectively. Further, all Q statistics were found to be positive while, GoF values for the model in each dataset were found to be large (All firms - 0.526, SME – 0.554 and Large firms – 0.513).

## 5 Discussion

This paper began with two objectives. The first objective was to identify the extent of the impact of environmental uncertainty on strategic alignment, and to determine its relative impact in light of the other antecedents. The second objective was to determine whether there were differences between small and medium, and large firms regarding relative impact of antecedents on strategic alignment. Table 3 summarizes the empirical results for the full sample as well as for each category of firms.

In meeting the first research objective, results indicated that environmental uncertainty significantly affects the planning connection. The literature argues that lack of information and unpredictability of the environmental variables (Fredrickson et al. 1984) adversely affect strategic planning (Johnson et al. 2005) and thus negatively affect strategic alignment (Sabherwal et al. 1994). However, the findings of this research reject this argument and provide empirical evidence that environmental uncertainty positively influences business-IT alignment, thus confirming the argument raised by Chan et al. (2006), who argue that given a lack of information, change and an unstable environment, businesses tend to rely more on IT and thus, environmental uncertainty has a positive influence on business-IT alignment.

In addition to environmental uncertainty, shared domain knowledge, relationship management and prior IS success were proposed to have a positive impact on strategic alignment. Results indicated that these three internal antecedents also significantly affect strategic alignment through two managerial practices. When the relative influence of the three internal antecedents were compared with the influence of the external antecedent – environmental uncertainty - the three internal antecedents were found to contribute more to the attainment of strategic alignment, showing a much higher path coefficient (see Figure 3). Thus, the findings suggest that business and IT executives should increase their knowledge of each other's domains, and that both business and IT executives need to make an effort to maintain a good relationship. Finally IT executives should be more proactive in terms of increasing the visibility of their success and making sure that they deliver on their promises.

With regard to the impact of the IT and business planning connection, and communications between IT and business strategic alignment, both demonstrated a significant influence on strategic alignment. When the effects of communication and planning connection are compared with each other, the results (in Figure 3) indicate that the positive effect of planning connection is much higher than that of communication. This suggests that the planning connection is more important than the level of communication between the executives in terms of achieving strategic alignment. Planning processes are more formal and consist of a set of phases and specific tasks within each phase (Newkirk et al. 2006); thus, taking action to align IT with business is much stronger in planning than in communication which is more embedded in relationships. The analysis strongly suggests that firms following an integrated planning approach improve their chances of achieving better business-IT alignment.

| Hypothesis | | All Firms | SME | Large |
|---|---|---|---|---|
| H1 | Communication between business and IT executives will positively influence the strategic alignment between business and IT | Supported | Not Supported | Supported |
| H2 | Planning connection between business and IT planning processes will positively influence the strategic alignment between business and IT | Supported | Supported | Supported |
| H3a | Shared business and IT domain knowledge will positively influence communication between business and IT executives | Supported | Not Supported | Supported |
| H3b | Shared business and IT domain knowledge will positively influence the connections between business and IT planning processes | Supported | Supported | Not Supported |
| H4a | Relationship management will positively influence the communication between business and IT executives | Supported | Supported | Not Supported |
| H4b | Relationship management will positively influence the connections between business and the IT planning processes | Supported | Not Supported | Supported |
| H5a | Prior IS success will positively influence the communication between business and IT executives | Supported | Supported | Not Supported |
| H5b | Prior IS success will positively influence the connections between business and IT planning processes | Supported | Supported | Supported |
| H6a | Environmental uncertainty will influence communication between business and IT executives | Not Supported | Not Supported | Not Supported |
| H6b | Environmental uncertainty will influence connection between the business and IT planning processes | Not Supported | Not Supported | Supported |

*Table 3. Summarized results of hypotheses testing.*
*The shaded cells indicate those hypotheses that were supported ($p<0.05$).*

The second objective was to provide empirical insights into the differences between SMEs and large firms regarding the relative impact of antecedents on the strategic alignment. As shown in Table 3, the effects of the antecedents and managerial practices vary between SME and large firms. Strategic direction in large firms is derived from years of accumulated data and experience making them more proactive to the environment (Parnell et al. 2012). Therefore, environmental uncertainty positively influences strategic alignment through the planning process. On the other hand, SMEs differ fundamentally from large firms, and these differences drive the way they deal with environmental uncertainty (Johnston et al. 2008). SMEs reaction to environmental uncertainty relies on the perception of owners/managers. They may not be as proactive as decision makers in large firms, resulting in less influence on strategic alignment.

Further, the results indicated that the relative influence of internal antecedents also varies between SMEs and large firms. For instance, in SMEs, relationship management significantly affects communication, but it does not affect the planning connection. With regard to large firms, the only significant effect is between relationship management and planning connection. Large firms are more likely to have a separate IT unit with its own goals and objectives. Therefore, in large firms it is important for IT and business executives to value each other's ideas during the planning process, and consider each other's goals and objectives, in order to achieve higher levels of planning integration. This result suggests that the relative impact of antecedents as well as the mechanisms used to attain strategic alignment vary by firm size.

## 6  Conclusion

Business-IT alignment has been one of the top concerns of practitioners and scholars. However, despite recognition of its positive effects on firm success, only few firms consider themselves to have achieved business-IT alignment (Luftman et al. 1999, Rosa 1998). This study has contributed to the literature on business-IT alignment in several ways. Firstly it has reviewed the alignment literature and developed a conceptual model to assess the influence of environmental uncertainty on business-IT alignment. Secondly, this study examined the impact of environmental uncertainty on alignment, and determined its relative impact in light of the other antecedents. The empirical results provide good overall support for the model and the arguments made in alignment studies. Further, the findings present a broader view of the alignment process, thus providing better guidance to executives for achieving and sustaining the business-IT alignment. Additionally, it has contributed to the field by showing that environmental uncertainty has a positive impact on business-IT alignment, but that internal antecedents have a greater impact on the attainment of strategic alignment. The research has also demonstrated that the mechanisms used to attain strategic alignment vary between SMEs and large firms.

One area for potential further research area would be to investigate the effect of each type of environmental uncertainty. Chan et al. (2006) demonstrated that some of the antecedents do not have significant effects on alignment for certain business strategies. Thus, another area for future research would be to investigate the effects of antecedents with regards to the business strategy.

## Appendix 1:

Consistency reliability, indicator reliability, and convergent validity of the constructs

| Item | All firms | | | SME | | | Large | | |
|---|---|---|---|---|---|---|---|---|---|
| | Loading | St.Err | p-value | Loading | St.Err | p-value | Loading | St.Err | p-value |
| **Strategic Alignment** | | | | | | | | | |
| | CR=0.880 / α=0.820 / AVE=0.646 | | | CR=0.893 / α=0.842 / AVE=0.677 | | | CR=0.853 / α=0.774 / AVE=0.594 | | |
| SA1 | 0.769 | 0.042 | 0.000 | 0.794 | 0.052 | 0.000 | 0.720 | 0.069 | 0.000 |
| SA2 | 0.810 | 0.037 | 0.000 | 0.846 | 0.049 | 0.000 | 0.742 | 0.074 | 0.000 |
| SA3 | 0.799 | 0.036 | 0.000 | 0.821 | 0.048 | 0.000 | 0.756 | 0.052 | 0.000 |
| SA4 | 0.838 | 0.024 | 0.000 | 0.829 | 0.045 | 0.000 | 0.857 | 0.026 | 0.000 |
| **Communication** | | | | | | | | | |
| | CR=0.862 / α=0.761 / AVE=0.676 | | | CR=0.862 / α=0.761 / AVE=0.675 | | | CR=0.862 / α=0.758 / AVE=0.676 | | |
| COM1 | 0.855 | 0.024 | 0.000 | 0.835 | 0.041 | 0.000 | 0.868 | 0.030 | 0.000 |
| COM2 | 0.826 | 0.035 | 0.000 | 0.802 | 0.070 | 0.000 | 0.853 | 0.026 | 0.000 |
| COM3 | 0.784 | 0.040 | 0.000 | 0.827 | 0.043 | 0.000 | 0.741 | 0.064 | 0.000 |
| **Planning Connection** | | | | | | | | | |
| | CR=0.912 / α=0.872 / AVE=0.723 | | | CR=0.920 / α=0.884 / AVE=0.742 | | | CR=0.905 / α=0.860 / AVE=0.705 | | |
| CON1 | 0.818 | 0.034 | 0.000 | 0.838 | 0.031 | 0.000 | 0.801 | 0.065 | 0.000 |
| CON2 | 0.841 | 0.026 | 0.000 | 0.845 | 0.046 | 0.000 | 0.839 | 0.033 | 0.000 |
| CON3 | 0.885 | 0.017 | 0.000 | 0.888 | 0.028 | 0.000 | 0.882 | 0.022 | 0.000 |
| CON4 | 0.856 | 0.023 | 0.000 | 0.874 | 0.035 | 0.000 | 0.834 | 0.036 | 0.000 |
| **Shared Domain Knowledge** | | | | | | | | | |
| | CR=0.906 / α=0.879 / AVE=0.579 | | | CR=0.897 / α=0.866 / AVE=0.555 | | | CR=0.913 / α=0.890 / AVE=0.601 | | |
| SDK1 | 0.738 | 0.039 | 0.000 | 0.725 | 0.070 | 0.000 | 0.747 | 0.047 | 0.000 |
| SDK2 | 0.755 | 0.040 | 0.000 | 0.741 | 0.062 | 0.000 | 0.778 | 0.057 | 0.000 |
| SDK3 | 0.804 | 0.024 | 0.000 | 0.784 | 0.042 | 0.000 | 0.818 | 0.036 | 0.000 |
| SDK4 | 0.695 | 0.042 | 0.000 | 0.611 | 0.081 | 0.000 | 0.757 | 0.046 | 0.000 |
| SDK5 | 0.797 | 0.032 | 0.000 | 0.801 | 0.053 | 0.000 | 0.791 | 0.042 | 0.000 |
| SDK6 | 0.767 | 0.039 | 0.000 | 0.768 | 0.068 | 0.000 | 0.764 | 0.051 | 0.000 |
| SDK7 | 0.764 | 0.042 | 0.000 | 0.768 | 0.069 | 0.000 | 0.768 | 0.044 | 0.000 |
| **Relationship Management** | | | | | | | | | |
| | CR=0.913 / α=0.872 / AVE=0.723 | | | CR=0.931 / α=0.901 / AVE=0.772 | | | CR=0.891 / α=0.837 / AVE=0.673 | | |
| RM1 | 0.853 | 0.026 | 0.000 | 0.872 | 0.037 | 0.000 | 0.831 | 0.039 | 0.000 |
| RM2 | 0.884 | 0.021 | 0.000 | 0.908 | 0.033 | 0.000 | 0.866 | 0.025 | 0.000 |
| RM3 | 0.848 | 0.038 | 0.000 | 0.899 | 0.030 | 0.000 | 0.792 | 0.074 | 0.000 |
| RM4 | 0.816 | 0.033 | 0.000 | 0.835 | 0.055 | 0.000 | 0.789 | 0.039 | 0.000 |
| **Prior IS Success** | | | | | | | | | |
| | CR=0.820 / α=0.676 / AVE=0.604 | | | CR=0.808 / α=0.648 / AVE=0.583 | | | CR=0.830 / α=0.697 / AVE=0.622 | | |
| PISS1 | 0.782 | 0.032 | 0.000 | 0.751 | 0.059 | 0.000 | 0.814 | 0.046 | 0.000 |
| PISS2 | 0.816 | 0.032 | 0.000 | 0.783 | 0.066 | 0.000 | 0.847 | 0.036 | 0.000 |
| PISS3 | 0.731 | 0.051 | 0.000 | 0.757 | 0.085 | 0.000 | 0.697 | 0.067 | 0.000 |

| Market Uncertainty | | | | | | | | |
|---|---|---|---|---|---|---|---|---|
| CR=0.866/ α=0.770/ AVE=0.683 | | | CR=0.875 / α=0.786 / AVE=0.699 | | | CR=0.860 / α=0.758 / AVE=0.672 | | |
| MU1  | 0.833 | 0.023 | 0.000 | 0.811 | 0.040 | 0.000 | 0.852 | 0.022 | 0.000 |
| MU12 | 0.815 | 0.032 | 0.000 | 0.869 | 0.036 | 0.000 | 0.773 | 0.052 | 0.000 |
| MU13 | 0.830 | 0.027 | 0.000 | 0.828 | 0.043 | 0.000 | 0.832 | 0.035 | 0.000 |
| Technological Uncertainty | | | | | | | | |
| CR=0.899/ α=0.732/ AVE=0.748 | | | CR=0.906 / α=0.844 / AVE=0.763 | | | CR=0.894 / α=0.823 / AVE=0.739 | | |
| TU1 | 0.833 | 0.021 | 0.000 | 0.852 | 0.023 | 0.000 | 0.814 | 0.035 | 0.000 |
| TU2 | 0.892 | 0.015 | 0.000 | 0.916 | 0.018 | 0.000 | 0.877 | 0.026 | 0.000 |
| TU3 | 0.869 | 0.019 | 0.000 | 0.852 | 0.031 | 0.000 | 0.885 | 0.021 | 0.000 |
| Regulatory Uncertainty | | | | | | | | |
| CR=0.878/ α=0.814/ AVE=0.644 | | | CR=0.915 / α=0.875 / AVE=0.729 | | | CR=0.891 / α=0.837 / AVE=0.562 | | |
| RU1 | 0.761 | 0.035 | 0.000 | 0.816 | 0.044 | 0.000 | 0.698 | 0.058 | 0.000 |
| RU2 | 0.890 | 0.015 | 0.000 | 0.915 | 0.020 | 0.000 | 0.861 | 0.029 | 0.000 |
| RU3 | 0.814 | 0.031 | 0.000 | 0.873 | 0.041 | 0.000 | 0.759 | 0.048 | 0.000 |
| RU4 | 0.737 | 0.031 | 0.000 | 0.805 | 0.051 | 0.000 | 0.666 | 0.076 | 0.000 |

## Appendix 2

Cross Loadings

|  | COM | CON | MU | PISS | RM | RU | SA | SDK | TU |
|---|---|---|---|---|---|---|---|---|---|
| COM1  | **0.855** | 0.517 | 0.177 | 0.477 | 0.506 | 0.176 | 0.448 | 0.495 | 0.209 |
| COM2  | **0.826** | 0.434 | 0.124 | 0.363 | 0.385 | 0.135 | 0.407 | 0.475 | 0.185 |
| COM3  | **0.784** | 0.462 | 0.180 | 0.392 | 0.430 | 0.226 | 0.434 | 0.390 | 0.248 |
| CON1  | 0.522 | **0.818** | 0.147 | 0.431 | 0.448 | 0.233 | 0.437 | 0.437 | 0.299 |
| CON2  | 0.521 | **0.841** | 0.142 | 0.500 | 0.484 | 0.219 | 0.512 | 0.444 | 0.263 |
| CON3  | 0.476 | **0.885** | 0.209 | 0.501 | 0.362 | 0.208 | 0.524 | 0.470 | 0.320 |
| CON4  | 0.439 | **0.856** | 0.146 | 0.464 | 0.391 | 0.189 | 0.528 | 0.419 | 0.222 |
| MU1   | 0.253 | 0.257 | **0.833** | 0.324 | 0.195 | 0.332 | 0.333 | 0.189 | 0.529 |
| MU2   | 0.070 | -0.007 | **0.815** | 0.020 | 0.095 | 0.299 | 0.012 | -0.015 | 0.405 |
| MU3   | 0.138 | 0.181 | **0.830** | 0.127 | 0.182 | 0.402 | 0.154 | 0.130 | 0.498 |
| PISS1 | 0.471 | 0.442 | 0.173 | **0.782** | 0.483 | 0.247 | 0.359 | 0.536 | 0.199 |
| PISS2 | 0.408 | 0.467 | 0.168 | **0.816** | 0.362 | 0.283 | 0.483 | 0.384 | 0.207 |
| PISS3 | 0.261 | 0.386 | 0.134 | **0.731** | 0.317 | 0.319 | 0.370 | 0.382 | 0.336 |
| RM1   | 0.425 | 0.440 | 0.135 | 0.417 | **0.853** | 0.275 | 0.392 | 0.490 | 0.257 |
| RM2   | 0.468 | 0.428 | 0.215 | 0.413 | **0.884** | 0.271 | 0.309 | 0.551 | 0.302 |
| RM3   | 0.479 | 0.379 | 0.184 | 0.435 | **0.848** | 0.225 | 0.323 | 0.491 | 0.233 |
| RM4   | 0.459 | 0.436 | 0.133 | 0.456 | **0.816** | 0.206 | 0.328 | 0.565 | 0.211 |
| RU1   | 0.185 | 0.227 | 0.274 | 0.333 | 0.225 | **0.761** | 0.257 | 0.186 | 0.392 |
| RU2   | 0.180 | 0.263 | 0.384 | 0.336 | 0.261 | **0.890** | 0.228 | 0.180 | 0.517 |
| RU3   | 0.201 | 0.138 | 0.351 | 0.251 | 0.253 | **0.814** | 0.134 | 0.062 | 0.452 |
| RU4   | 0.133 | 0.161 | 0.332 | 0.221 | 0.176 | **0.737** | 0.106 | 0.000 | 0.390 |
| SA1   | 0.319 | 0.398 | 0.188 | 0.381 | 0.252 | 0.181 | **0.769** | 0.401 | 0.259 |
| SA2   | 0.352 | 0.435 | 0.093 | 0.466 | 0.300 | 0.125 | **0.810** | 0.392 | 0.149 |
| SA3   | 0.460 | 0.474 | 0.238 | 0.372 | 0.372 | 0.186 | **0.799** | 0.422 | 0.264 |
| SA4   | 0.512 | 0.560 | 0.179 | 0.454 | 0.338 | 0.230 | **0.838** | 0.469 | 0.315 |
| SDK1  | 0.492 | 0.428 | 0.094 | 0.438 | 0.455 | -0.027 | 0.439 | **0.738** | 0.161 |
| SDK2  | 0.394 | 0.333 | 0.003 | 0.376 | 0.369 | 0.003 | 0.322 | **0.755** | 0.090 |
| SDK3  | 0.443 | 0.497 | 0.175 | 0.530 | 0.458 | 0.158 | 0.472 | **0.804** | 0.246 |
| SDK4  | 0.335 | 0.313 | 0.120 | 0.378 | 0.409 | 0.221 | 0.297 | **0.695** | 0.183 |
| SDK5  | 0.420 | 0.367 | 0.174 | 0.409 | 0.496 | 0.074 | 0.427 | **0.797** | 0.166 |
| SDK6  | 0.398 | 0.366 | 0.017 | 0.380 | 0.537 | 0.162 | 0.366 | **0.767** | 0.133 |
| SDK7  | 0.431 | 0.428 | 0.116 | 0.461 | 0.554 | 0.170 | 0.441 | **0.764** | 0.159 |
| TU1   | 0.289 | 0.329 | 0.440 | 0.342 | 0.320 | 0.397 | 0.385 | 0.245 | **0.833** |
| TU2   | 0.189 | 0.242 | 0.497 | 0.209 | 0.221 | 0.460 | 0.160 | 0.115 | **0.892** |

| TU3 | 0.200 | 0.273 | 0.574 | 0.244 | 0.228 | 0.560 | 0.268 | 0.202 | **0.869** |

*COM: Communication, CON: Planning Connection, EU: Environmental Uncertainty, MU: Market Uncertainty, PISS: Prior IS Success, RM: Relationship Management, RU: Regulatory Uncertainty, SA: Strategic Alignment, SDK: Shared Domain Knowledge, TU: Technology Uncertainty.*

## Appendix 3

Global goodness-of-fit (GoF) statistics

|  | All firms | | SME | | Large | |
| --- | --- | --- | --- | --- | --- | --- |
| Endogenous construct | AVE | $R^2$ | AVE | $R^2$ | AVE | $R^2$ |
| Strategic alignment | 0.646 | 0.398 | 0.677 | 0.409 | 0.594 | 0.385 |
| Communication | 0.676 | 0.403 | 0.675 | 0.464 | 0.676 | 0.371 |
| Planning connection | 0.723 | 0.407 | 0.742 | 0.398 | 0.705 | 0.442 |
| Average | 0.682 | 0.405 | 0.698 | 0.424 | 0.658 | 0.399 |
| GoF | 0.526 | | 0.544 | | 0.513 | |

## Appendix 4

Predictive relevance statistics

|  | All firms | SME | Large |
| --- | --- | --- | --- |
| Endogenous construct | $Q^2$ | $Q^2$ | $Q^2$ |
| Strategic alignment | 0.272 | 0.291 | 0.227 |
| Communication | 0.261 | 0.274 | 0.237 |
| Planning connection | 0.282 | 0.269 | 0.281 |